\documentclass[useAMS,usenatbib]{mn2e}

\usepackage{graphicx}

\usepackage{color}

\usepackage{hyperref}

\newcommand{\rmvec}[1]{\ensuremath{\bmath{#1}}}
\newcommand{\tens}[1]{\mathbfss{#1}}

\newenvironment{textchange}{}%
{}
\newcommand{\textch}[1]{{#1}}

\title[Photometric mode identification I. -- Dynamic Eclipse Mapping]{Photometric mode identification methods \\
       of nonradial pulsations in eclipsing binaries \\
       I. --  Dynamic Eclipse Mapping}
\author[I. B. B\'{\i}r\'{o} and J. Nuspl]{I. B. B\'{\i}r\'{o}$^{1}$ 
\thanks{barna\@electra.bajaobs.hu} and J. Nuspl$^{2}$\thanks{nuspl@konkoly.hu}\\
$^{1}$ Baja Astronomical Observatory, Szegedi \'{u}t, Kt. 766, Baja 6500, Hungary\\
$^{2}$ Konkoly Observatory of the Hungarian Academy of Sciences, Konkoly--Thege \'{u}t 13--17, Budapest 1121, Hungary}

\begin{document}

\date{Accepted 2011 January 21.  Received 2011 January 19; in original form 2010 September 16.}

\pagerange{\pageref{firstpage}--\pageref{lastpage}} \pubyear{2010}

\maketitle

\label{firstpage}

\begin{abstract}
   We present \textbf{the} Dynamic Eclipse Mapping method designed specifically to reconstruct
the surface intensity patterns of non-radial stellar oscillations on components of eclipsing binaries. The method needs a geometric model of the binary, accepts the light curve and the detected pulsation frequencies on input, and on output yields estimates of the pulsation patterns, in form of images -- thus allowing a direct identification of the surface mode numbers $(\ell,m)$. 
Since it has minimal modelling requirements and can operate on photometric observations in arbitrary wavelength bands, Dynamic Eclipse Mapping is well suited to analyze the wide-band time series collected by space observatories.

  We have investigated the performance and the limitations of the method through extensive numerical tests on simulated data, in which \textch{almost all photometrically detectable modes with a latitudinal complexity $\ell-|m|\le4$ were properly restored.}
  The method is \textch{able by its nature} to simultaneously reconstruct \textch{multimode pulsations} from data covering a sufficient number of eclipses, as well as pulsations on components with tilted rotation axis of known direction. It can also be applied in principle to isolate the contribution of hidden modes \textch{from the light curve}.

   Sensitivity tests show that moderate errors in the geometric parameters and the assumed limb darkening can be partially tolerated by the inversion, in the sense that the lower degree modes are still recoverable. 
  Tidally induced or mutually resonant pulsations, however, are an obstacle that neither the eclipse mapping, nor any other inversion technique can ever surpass.
    
  We conclude that, with reasonable assumptions, Dynamic Eclipse Mapping could be a powerful tool for mode identification, especially in moderately close eclipsing binary systems, where the \textch{pulsating component is not seriously affected} by tidal interactions \textch{so that the pulsations are intrinsic to them, and not a consequence of the binarity.}
\end{abstract}


\section{Introduction}

In recent years an increasing number of eclipsing binary systems have been discovered
to harbor pulsating components. The latest comprehensive catalog 
\citep{zhou.2010:ebnrp-list} contains $99$ such objects, discovered almost exclusively from the ground.
The majority ($70$ in total) show delta Scuti type pulsations, followed by $8$ sdB and $5$ $\beta$~Cephei type pulsators as the second and third most frequent types. Even more are expected to be discovered from the current space-based missions \textit{MOST}, \textit{CoRoT}, and \textit{Kepler}.
There are \textch{clear} indications that at least some of the modes are non-radial (as expected for such pulsators). For example, in systems like RZ~Cas \citep{rzcas:nrp} or Y~Cam \citep{ycam:nrp}, the pulsations show amplitude and phase modulations during the eclipse phases, as a consequence of the symmetry violation in the surface flux integral due to the occultation.

This opens a new avenue of opportunities for asteroseismic investigations.
As has been shown by numerous examples, there is a major difficulty in single stars that renders the mode identification a state of the art procedure: the inability to resolve their surface. The observables (fluxes, spectral line features) are weighted integrals of the local quantities over the visible stellar disc, and therefore show only weak dependence on the pulsation modes. 
All photometric and spectroscopic mode identification methods
(for the most widely used ones, see \citealp{watson88:pm-id,balona-evers02:pm-id};
\citealp{briquet-aerts03:sp-moment};  
 \citealp{zima06:sp-fourier}
) 
must employ detailed models of the internal structure, atmosphere and pulsation, in order to overcome the problem of the low sensitivity. Accurate observational data and model parameters make a nearly unambiguous mode identification possible, with $\ell$ and $m$ determined to an accuracy of $\pm 1$ or better. Unfortunately, single stars rarely have well-determined parameters. 
The more general approach of Doppler Imaging
\citep{berdyugina.etal.03:di.nrp,kochukhov.04:di.nrp}, a remarkable 
technique for single stars, needs less sophisticated models because it aims at an image-like reconstruction of the surface patterns, and is therefore less sensitive to errors in the stellar parameters \citep{kochukhov.04:di.nrp}; however, it is
only applicable for rapid rotators, and the solutions still suffer from \textch{ambiguities} \citep[see, e.g.,][]{berdyugina.etal.03:di.papII}.

In contrast, a pulsating star in an eclipsing binary offers
at least two advantages over the single star scenario.
First, binarity enables a precise determination of the fundamental stellar parameters.
Second, the eclipses -- the mutual occultations of the stars -- 
implicitly provide a surface sampling: the shadow of one component
literally sweeps across the surface of the other. This purely geometric
phenomenon convolves the brightness distribution of the stellar surfaces into the variation of the integrated flux -- that is, the light curve.
Various inversion techniques may be used to recover the surface brightness structure from the light curve. They mainly differ in the amount of \textit{a priori} assumptions about the surface pattern. 
The most common assumption, that the pulsations can be described by spherical harmonics, was employed by \citet{gametal03:spatial-filtering}, using the concept of spatial filtering \citep{nather.robinson.74:sf} to identify non-radial modes in the interacting Algol-type binary RZ~Cas. 
Although the modes could not be unambiguously identified -- partly due to the complicating nature of the mass transfer between the components and partly because their approach was not suited very well to multiperiodic oscillations -- their study was the first to demonstrate the potential of the approach.
\begin{textchange}
Recently, \citet{ycam:modeint}, based partially on the above method, made a preliminary mode identification for the 8 modes discovered in Y~Cam, with a similar ambiguity in the mode numbers.
\end{textchange}

More generally, Eclipse Mapping methods can be used to invert \textch{photometric time series} into an instant image of the surface intensity distribution. When applied to high precision photometric data with appropriate \textch{temporal} resolution, they can discern far more detailed surface structure than the conventional methods. 
Eclipse Mapping techniques have already been \textch{used} to reconstruct static intensity  patterns in a variety of eclipsing binaries (accretion discs in cataclysmic binaries, close binaries with spotted members, and even contact binaries).
An immediate opportunity is then to map the non-radial oscillation patterns in eclipsing binaries, which would allow a \textch{direct approach} to the mode identification.
As with Doppler Imaging, Eclipse Mapping needs only simple models with a few parameters, which, moreover, can be more easily determined in eclipsing binaries. 
Therefore, asteroseismology could in principle be made much easier for the pulsators in eclipsing binaries.
In addition, the \textch{commonly employed} approximation of non-radial pulsations with single spherical harmonics becomes questionable for moderate rotation speeds already
\citep{aerts.eyer.00:modeint-lpv}, and is certainly invalid in rapid rotators  \citep{lignieres.etal.06:papI,reese.etal.06:papII}. 
Many single delta Scuti stars are quite rapid rotators
\citep{rodriguez.etal.00:dsct-cat}, and rotation speeds of the same order are expected in binary systems with synchronized orbits. Obviously, Eclipse Mapping is a more realistic approach for such cases than fitting patterns of some analytic form.

In the present paper we describe a variant of Eclipse Mapping method, \textit{Dynamic Eclipse Mapping}, designed to reconstruct the surface intensity patterns of nonradial  pulsations in eclipsing binary stars. We show that, under plausible circumstances, its application makes the mode identification possible \textch{in a large variety of eclipsing binary scenarios}.
%
%
%
\section{Dynamic Eclipse Mapping\label{sec:em-desc}}
%
%
\subsection{Eclipse Mapping of surface patterns}
%
The method of Eclipse Mapping (EM) was originally conceived to reconstruct the intensity 
distribution of radiation from accretion discs in eclipsing cataclysmic variables (CVs), with special 
emphasis on their radial temperature profile as a valuable diagnostic tool 
\citep{em:horne83phd,horne.85:em}. Its overwhelming success in different CV 
scenarios (see \citealp{baptista.04} for a summary) has then inspired widespread 
applications in other, less exotic eclipsing systems too, like mapping surface 
brightness inhomogeneities (spots) in late-type close binaries \citep[e.g. 
XY~UMa,][]{collcam.97:em.spot} or in W~UMa stars \citep[e.g. 
VW~Cep,][]{hendry.92:em.wuma}. 
A simplified EM was also used to estimate the surface temperature map of an exoplanet occulted by its host star \citep{knutsonetal06:em-exo}.

\begin{textchange}
The basic idea is that the eclipse acts as a surface sampler that convolves the surface brightness distribution into the light curve. Eclipse Mapping does the inverse,
it performs a deconvolution of the light curve into a surface pattern. The latter is treated as an \textit{image}%
\footnote{Throughout the paper, the terms \textit{image}, \textit{pattern} and \textit{map} are all used to refer to the same discretized surface brightness distribution.}%
composed of pixels on an appropriate surface grid. 

No analytic assumptions are thus made on the surface brightness distribution, which allows the method to be maximally free of stellar interior and atmospheric models,
and makes it particularly suited to confront model predictions with observations in a maximally unbiased manner. All it requires is a proper modelling of the eclipses and some basic atmosphere parameters.
But because the convolution of a two-dimensional distribution into a one-dimensional time series implies a considerable loss of information, the inversion is ill-posed. The regularization is made by introducing additional, \textit{a priori} information about the solution in the form of a \textit{regularization functional}, $\cal S(\rmvec{f})$, which measures some desired property of the image $\rmvec{f}$. The solution is obtained by maximizing this functional with respect to the elements of $\rmvec{f}$ and subject to the constraint of fitting the data at a prescribed level. The latter is usually measured by the chi-squared function.

Classical Eclipse Mapping uses the information entropy of the image as the regularization function,
\begin{equation}
\textstyle
{\cal S}(\rmvec{f}, \rmvec{A}) 
= - \sum_{k=1}^{N} 
  \bigl( \mathrm{f}_k - A_k - \mathrm{f}_k \ln ({\mathrm{f}_k}/{A_k}) \bigr)
\label{eq:cross-entropy}
\end{equation}
\citep{horne.85:em,shore.johnson.80:ent.axioms,skilling.88:shannon-jaynes}, which measures the negative of the information content of the image \textit{vector} $\rmvec{f}$ against that of a \textit{reference map} $\rmvec{A}$. The latter may be used to implement additional, user-defined preferences, or may be just a uniform map, scaled so that its total flux equals that of $\rmvec{f}$. The solution is then the image with the least structure (relative to the reference map) that can explain the observed data.

Other regularization methods may be employed equally well; \citet{kaipio.somersalo.05} provide a thorough presentation of possibilities, while \citet{craig.brown.86:inv.astr} focuses on inverse problems found specifically in astronomy.
In particular, the choice of the regularization functional $\cal S$ depends on the type of the problem. For example, the information entropy is a good choice for positive images with uncorrelated pixels. Physical maps are seldom uncorrelated; but the correlation can be easily accounted for via the reference map (see Sec.~\ref{sec:refmodel}).
Alternatively, the Tikhonov functional ${\cal{S}}(\rmvec{f})=\left\|\rmvec{f}\right\|^2$ \citep{tikhonov63} and its derivatives, which measure the smoothness of the image, are also used in some cases, e.g., \citet{piskunov.et.al.90}.
The optimization can be accomplished by standard methods. Most commonly it is transformed to a series of unconstrained optimizations by the method of Lagrange multipliers, where the multiplier plays the role of the regularization parameter, which is tuned between the optimizations until the desired level of data fitting is achieved. \citet{mem-genalg} give a sophisticated algorithm which accomplishes the two tasks in parallel, resulting in a robust and fast code. It is the algorithm that we have adopted in our implementation.
\end{textchange}
\begin{textchange}
\subsection{Eclipse Mapping of pulsation patterns}
\label{sec:method}

Previous applications of the Eclipse Mapping reconstructed surface structures that were \textit{static} in time, or at least implicitly considered static during the data collection period. In principle only the underlying model needs to be changed in order to handle the time-dependent patterns. The time dependence, however, has implications on other aspects too. 

\subsubsection{Assumptions}
\label{sec:assumptions}

The basic requirement of Eclipse Mapping is that the geometric configuration of the binary must be known. Pulsations, however, cause a periodic variation of the stellar shape, hence also in the local gravity and other atmospheric conditions. Their inclusion
would require not only detailed atmospheric model (see \citealp{buta.smith.79,townsend.97}), but also the knowledge of the pulsation modes themselves. Obviously, this is not feasible. 
Fortunately, for small amplitude pulsations these effects can be neglected. Indeed, for delta Scuti stars, the observed amplitudes in radial velocities, combined with typical pulsation periods, yield negligible displacements compared to the stellar radius. 
Variations of the surface normal could still be significant for high surface degree $\ell$, but, according to \citet[][Fig.~6.4.]{pulsbook}, photometrically detectable modes are limited by cancellation effect to $\ell=4$%
.
Although the eclipses break the symmetry of the disc integration and may thus amplify some modes of even higher degree, those modes would still need to be detected outside the eclipses to have their frequency determined.
\end{textchange}
Therefore, these variations can be safely neglected in our case, and the star is
considered a rigid body of known shape.

At the present stage we use spherical stars on circular orbits as the model of the binary. Limb darkening is taken into account (but its small variations with the local atmospheric conditions are again neglected), but we make no further assumptions about the stellar atmosphere, using fluxes of arbitrary spectral range.

\begin{textchange}
For real, not-so detached systems, a proper account of the secondary's distorted shape may have to be made; although we note that its projection on the sky during the eclipses is still close to circular, so we limited ourselves to spherical secondary for the evaluation of the method's capabilities.

More stringent conditions are imposed on the pulsating star, for which rotation is a complicating factor. Besides its physical influence on the pulsations, rotation has a simple geometric effect that the frequency detected in the observer's frame will differ from the physical frequency in the co-rotating frame, according to the relation
\begin{equation}
   \omega_{\rmn{obs}} = \omega_{\rmn{surf}} + m \Omega_{\rmn{rot}}, \nonumber
\end{equation}
where $\Omega_{\rmn{rot}}$ is the angular rotation velocity, and $m$ is the azimuthal order of the mode. 
Even if the rotation velocity were available, the physical frequency is unknown, because $m$ is also unknown, being one of the parameters sought by our analysis. 
The time dependence of the patterns on the rotating stellar surface being thus unavailable, they cannot be mapped! What is known to us is the frequency of the patterns as seen on the visible, non-rotating stellar hemisphere. Therefore we have to map those patterns, and infer the whole-surface patterns from them. Obviously, this requires axial symmetry of both the stellar surface -- in order for the shape of the disc to be constant in time -- and of the pulsation patterns themselves -- to assure that the surface and sky-projected patterns are equivalent, i.e., the pulsation amplitudes and phases seen on the visible stellar hemisphere correspond to those of the intrinsic pattern.

Unfortunately, the above restriction excludes oscillations of a tidally distorted star to be mapped with this technique, because both the shape of the projected disc and the oscillation amplitudes do vary in time. Oblique pulsators, characteristic of roAp stars, would also be difficult to handle, the amplitudes being modulated by the stellar rotation \citep{kurtz.82:roAp}. In this respect, wider systems, where binarity and pulsation are a mere coincidence, are the preferred targets for the Dynamic Eclipse Mapping.
\end{textchange}

\begin{textchange}
\subsubsection{Statement of the problem}

We may turn now to the mathematical formulation. For small amplitudes, each pulsation mode can be written as a sinusoidally oscillating perturbation to the equilibrium intensity distribution of the star, so the time-dependent (visible) surface intensity pattern is the superposition of a static equilibrium map, $\rmvec{f}^{(0)}$, and $P$ sinusoidally oscillating patterns:
\begin{equation}
 \rmvec{f}(t) = \rmvec{f}^{(0)} + 
  \textstyle\sum_{\nu=1}^{P} 
  \left[ \rmvec{C}^{(\nu)} \, \cos(\omega_\nu t) + \rmvec{S}^{(\nu)} \, \sin(\omega_\nu t)
  \right],
 \label{eq:surf-pattern}
\end{equation}
where $\omega_\nu$ is the frequency of mode $\nu$, and the 'cosine' and 'sine' maps $C^{(\nu)}_k=A^{(\nu)}_k\,\cos F^{(\nu)}_k$ 
and 
$S^{(\nu)}_k=A^{(\nu)}_k\,\sin F^{(\nu)}_k$
have been used rather than the amplitude and initial phase maps $(\rmvec{A},\rmvec{F})$ 
because they have the same units, of intensity.

In the linear adiabatic  approximation and for slow rotation, the pulsation patterns would be described in terms  of spherical harmonics $Y_{\ell}^{m}$. In the general approach, however, they are just pairs of images to be reconstructed. 

The dataset of integrated fluxes at $M$ discrete moments is obtained by convolving the image $\rmvec{f}$, composed of $N$ pixels, with an \textit{occultation kernel} $\tens{K}$, an $M\times N$ matrix, each row of which contains the pixels' contribution factors to the specific datapoint. These factors are composed of the projected area of the visible pixel portion (including the foreshortening factor), the limb darkening, and eventually other known factors incorporated into the model of the binary (proximity effects, for example). 
With the image~(\ref{eq:surf-pattern}), and denoting the times as 
$t_\phi (\phi=1\dots M)$, the resulting synthetic dataset is
\begin{eqnarray}
y_\phi
 & = &  
  \textstyle \sum_{k=1}^{N} K_{\phi k} f_{k}(t_\phi) \nonumber \\
 & = & 
\textstyle
 \sum_{k=1}^{N} K_{\phi k} f^{(0)}_k + \nonumber\\
     &  & +  \textstyle 
         \sum_{\nu=1}^{P} 
         \sum_{k=1}^{N} \left[ 
            {K_{[\rmn{c}]}}^{(\nu)}_{\phi k}\,C^{(\nu)}_k +
            {K_{[\rmn{s}]}}^{(\nu)}_{\phi k}\,S^{(\nu)}_k
                        \right],
\label{eq:convpuls}
\end{eqnarray}
where we introduced the 'cosine' and 'sine' kernels
\begin{equation}
\begin{array}{lcl}
{K_{[\rmn{c}]}}^{(\nu)}_{\phi k} 
& = & K_{\phi k}\,\cos\,(\omega_\nu t_\phi),\\
{K_{[\rmn{s}]}}^{(\nu)}_{\phi k} 
& = & K_{\phi k}\,\sin\,(\omega_\nu t_\phi),
\end{array}
\label{eq:newkern}
\end{equation}
obtainable from scaling the rows of the base kernel $\tens{K}$ with the corresponding cosine and sine time factors.
With this notation, Equation (\ref{eq:convpuls}) can be written in a compact vectorial-tensorial form as:
\begin{equation}
\rmvec{y} = 
     \tens{K} \cdot \rmvec{f}^{(0)} +
     \sum_{\nu=1}^{P} \left( 
       {\tens{K}_{[\mathrm{c}]}}^{(\nu)} \cdot \rmvec{C}^{(\nu)} +
       {\tens{K}_{[\mathrm{s}]}}^{(\nu)} \cdot \rmvec{S}^{(\nu)}
     \right).
\end{equation}
Thus the known time dependence is transferred into the kernels, so that the model parameters are: i) one map for the static pattern, 
$\rmvec{f}^{(0)}$, and ii) a pair of cosine and sine maps
$(\,\rmvec{C}^{(\nu)}, \rmvec{S}^{(\nu)}\,)$ for each pulsation mode,
making together $2 P + 1$ independent maps in total. 

The task is to estimate the $2P+1$ maps for a known kernel 
$\tens{K}$, and a given set of frequencies $\omega_\nu$ $(\nu=1\dots P)$ as well as the light curve $\rmvec{d}$. The kernel is provided by the model for the binary, while the frequencies and the light curve are the observed data.

\subsubsection{Regularization}
Because the pulsation patterns $\rmvec{C}$ and $\rmvec{S}$ are not strictly positive but may contain values of either sign, the entropy expression (\ref{eq:cross-entropy}) 
is not valid for them (it can still be used for the static pattern $\rmvec{f}^{(0)}$, though). For $\rmvec{C}$ and $\rmvec{S}$ we selected a simple quadratic regularization function, of form
\begin{equation}
{\cal S}(\rmvec{f},\rmvec{A}) = - \| \rmvec{f} - \rmvec{A} \|^2 = - \sum_k (f_k - A_k)^2,
\label{eq:quad-reg}
\end{equation}
which also includes a reference map $\rmvec{A}$. 

This function is in fact a generalized Tikhonov functional, although we arrived at it through the statistical approach to inverse problems \citep{kaipio.somersalo.05}, where the regularization function can be interpreted as the logarithm of an \textit{a priori} probability distribution function (pdf) of the parameters. The above quadratic expression may be recognized as the logarithm of a joint Gaussian pdf, apart from a constant.
The Gaussian is considered the most 'generic' pdf (or most noncommittal with regard to missing information) when only the first two moments of the parameter are available
\citep[][Chap.~7 and 11]{jaynes.book}. In our case, the first moment -- the expectation value -- is given by the reference map $\rmvec{A}$, while the second moment -- the 'spread' -- plays the role of the regularization parameter. 
(Similarly, the entropy expression~(\ref{eq:cross-entropy}) is the logarithm of a Poissonian pdf, with mean pixel values given in the reference map $\rmvec{A}$, e.g, \citealp{skilling.98:massinf}.)

The entropy expression~(\ref{eq:cross-entropy}) for the equilibrium intensity map and the quadratic functionals~(\ref{eq:quad-reg}) for the pulsation patterns must then be 
\textit{simultaneously} optimized, subject to a common constraint 
$\chi^2= \chi_{\rmn{aim}}^2$. 
A pair of maps belonging to one mode may be handled as a single entity, their regularization functions are simply summed up, so we have $P+1$ objective functions to optimize, each on its subset of variables. This multiobjective optimization problem
can be transformed with the simple weighting method to a single optimization of their weighted sum $\sum_{\nu=1}^{P+1} w_\nu S_\nu$, with the weights determined by the desideratum that all maps should have about the same smoothness (as measured by the value of the regularization function).
Because the objectives only interfere indirectly with each other via the common data fitting constraint (they have disjunct subsets of variables), the \textit{ideal point method} can be employed for computing the weights (see \citealp{liu.etal:mop.book}, for the aforementioned methods).
This involves performing $P+1$ optimizations first, with only one objective active at a time, but with all maps being fitted. A higher range of achieved values for an objective means more room to make the corresponding maps smoother, and translates to a higher weight in the final optimization run.

We have found the Variable Chi Algorithm described by \citet{baptista.steiner.93:vca} quite useful in setting up a reliable criterion for data fitting. They introduced a so-called $R$-statistics measuring the correlations of the neighboring residuals. It was shown that the $R$ and $\chi^2$ statistics are proportional to each other, and setting  a goal value for $R$ as $R_{\rmn{aim}}\sim0.5-1$ provides a more data-independent fitting criterion for $\chi^2$ -- automatic noise scaling, in fact.

The stopping criterion of the iterations was the same as in \citet{mem-genalg}. After reaching a good fitting to the data, the iterations were continued until a TEST value, which measures the parallelism between the gradients of ${\cal S}(\rmvec{f})$ and $C(\rmvec{f},\rmvec{d})$, decreased below a certain value \textch{in all subsets}. Being half the sine of the angle between the two involved vectors (in terms of the scalar product), TEST takes values from $0$ to $0.5$. The algorithm could routinely reach below TEST=$0.01$, which we chose as the stopping value.


\subsubsection{The reference map}
\label{sec:refmodel}

The role of the reference map $\rmvec{A}$ appearing in the regularization functionals 
is to allow the introduction of additional user preferences about the solution.
In highly ill-posed inverse problems like Eclipse Mapping, where every bit of a priori knowledge is important, it may have a determining role in obtaining a proper solution.

The regularization functionals have their global maximum at the location of the reference map, which therefore is the default solution in the absence of observational constraints. Regular solutions will also be as close to it as allowed by the constraints. 
Updating the reference map from the instantaneous solution during the iterative solving procedure is a common technique to make it control the image property to be measured by the regularization function, which is hence optimized in the solution (see \citealp{horne.85:em} and \citealp{bobinger.etal.99:ddem} for examples in accretion disc reconstructions).
Alternatively, it may be set to a uniform map if no additional preferences exist; in this case we obtain a 'most uniform', but spatially uncorrelated, solution: permuting the pixels will not change the value of the regularization function at all. This is better than nothing; but most problems do have some symmetry, and employing it improves the solution. 

The pulsation patterns of a rotationally symmetric star, when described in a spherical coordinate system tied to the rotation axis, obey a kind of axial symmetry, in that the local amplitude only depends on the latitude, while the local phase only varies with the longitude. This holds not only for spherical harmonics, but for all pulsations of tidally \textit{un}distorted stars -- including fast rotators, for example.
Consequently all the 'cosine' and 'sine' maps are expected to have the form 
$C(\theta,\varphi) = A(\theta)\cos F(\varphi)$, and 
$S(\theta,\varphi) = A(\theta)\sin F(\varphi)$, 
where $\theta$ and $\varphi$ are the co-latitude and longitude, respectively.
On a uniform spherical grid, therefore, we have a discrete number of amplitudes $A_k=A(\theta_k) \enspace (k=1\dots N_\theta)$ 
and initial phases 
$F_l=F(\varphi_l) \enspace (l=1\dots N_\varphi)$,
that completely describe the maps $\rmvec{C}$ and $\rmvec{S}$ on that same grid:
\begin{equation}
\begin{array}{lcl}
C_{kl} & = & A_k \, \cos F_l,\\
S_{kl} & = & A_k \, \sin F_l.
\end{array}
\label{eq:ref-model}
\end{equation}

According to this a priori expectation, we set up the following updating scheme for the reference maps.
The algorithm starts with computing flat-valued images by linear least-squares fitting to the data. The reference maps are initialized with the same images.
After the end of each iteration the latest $(\rmvec{C},\rmvec{S})$ solutions are interpolated onto a uniform spherical grid around the rotation axis, of appropriate resolution \mbox{$N_\theta \times N_\varphi$}. The non-linear model (\ref{eq:ref-model}) 
with the parameters $A_1\dots A_{N_\theta},F_1\dots F_{N_\varphi}$ is fitted to the maps, by a standard Levenberg--Marquardt algorithm. Spatial correlations are then taken into account by smoothing the resulting discretized amplitude and phase profiles with a Gaussian of a user-supplied angular correlation length. From the smoothed profiles, new $(\rmvec{C},\rmvec{S})$ maps are computed and interpolated back to the reconstruction grid, giving the reference maps for the next iteration.
For the equilibrium map $\rmvec{f}^{(0)}$, only a surface Gaussian smearing is applied; this takes spatial correlations into account (i.e., to yield a smooth image).

A few safety measures were needed for proper operation of the above procedure. These include a careful estimation of starting parameter values for the non-linear model fitting algorithm, as well as making the smoothing wraparound for the phase profiles (running along the longitude) and mirrored at the poles for the amplitude profiles, in order to ensure continuity over the stellar surface. 

The same procedure is used when amplitude and phase profiles are derived from the final solution for the purpose of mode identification, with the difference that the profiles are not smoothed.

The above scheme only works for a known rotation axis. Stars in eclipsing binaries are generally assumed to have the rotation axis perpendicular to the orbital plane. This assumption is expected to hold for close binaries, where tidal interactions tend to bring the spin axes parallel to the orbital axis, in addition to circularizing the orbits. For wider systems, which are of primary interest for us (Sec.~\ref{sec:assumptions}), it may not be true, as demonstrated by the recent discovery of very oblique rotation axes (being almost in the orbital plane) in the eccentric system DI~Her \citep{albrecht.etal.10:di.her}.
If the rotation axis is unknown, then only the assumption of smoothness can be used, leading to a 'smoothest' solution, which will for certain be inferior to the regular solution, and may be insufficient for mode identification. Assuming a wrong axis would do even worse, though.

We note that the seemingly ad-hoc manipulation of the reference map may be justified within the frame of the Bayesian approach: each iteration is a prior-to-posterior processing step, and the reference map updating only prepares the prior for the next iteration from the posterior of the last iteration, by propagating only those structures that are consistent with the a priori expectations, and ignoring everything else.
\end{textchange}

\subsection{The role of the eclipse geometry}
\label{sec:egeom}

The amount of recoverable surface structure information depends primarily on how the surface is sampled by the eclipses. Uneclipsed regions, for example, cannot be reconstructed; but we cannot ignore them either because they contribute to the integrated flux. Therefore pixels within these regions are fitted, but not 'regularized'. Being spatially unresolvable, the uneclipsed region is replaced by a single parameter, a \textit{virtual pixel}, with its contribution factor equivalent to that of the whole region. The value of this pixel is simply copied to the virtual pixel of the corresponding reference map during the updating scheme, therefore it gives zero net contribution to the regularization function, but still counts to the goodness-of-fit. This technique also prevents an unnecessary proliferation of the parameters.
Each map has its own virtual pixel of this kind; the one pertaining to the static intensity map takes in addition the duty of fitting any other uneclipsed flux ('third light') in the system.

\begin{figure}
\begin{center}
 \includegraphics[width=1.5in]{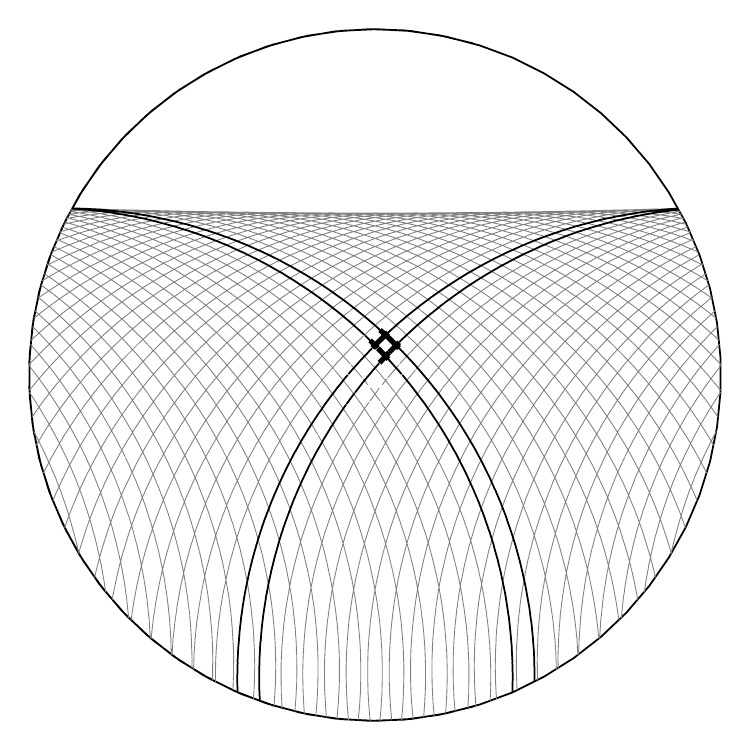}  
 \includegraphics[width=1.5in]{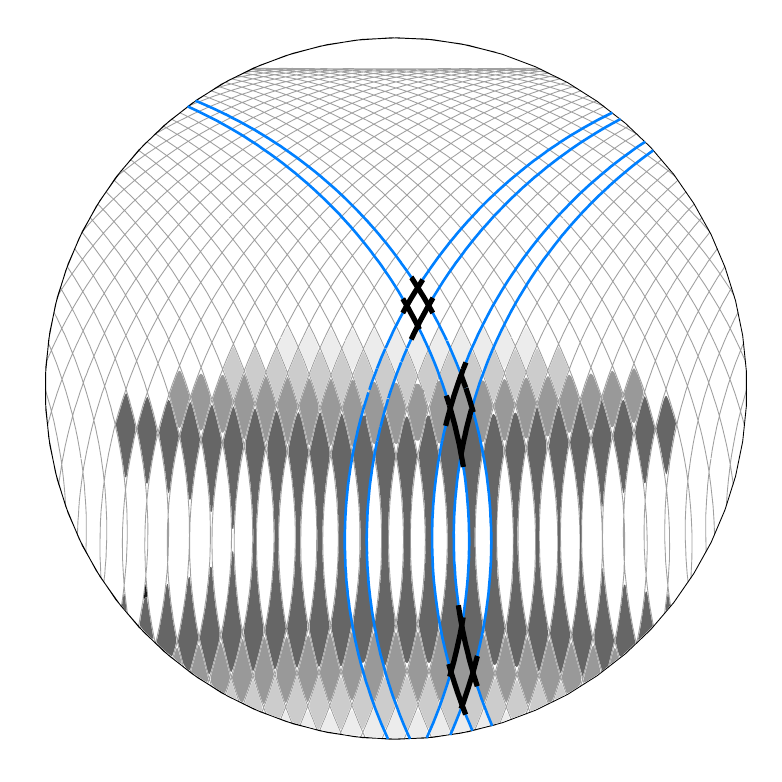}  
 \caption{The sampling grids of the primary's sky-projected disc for a binary system with fractional radii $R_1=0.195$ and $R_2=0.265$, seen under inclinations of $80\degr$ (\textit{left panel}) and $85\degr$ (\textit{right panel}), respectively. The secondary sweeps across the primary in the horizontal direction. The white upper areas are never eclipsed. Each shade of the grey-coloured pixels marks  one specific chain of equivalent pixel pairs on the right panel. One such pair is highlighted with thick black lines. Both panels also show a pixel with no equivalent pair.
\label{fig:egeom}
}
\end{center}
\end{figure}

Although systems with total eclipses seem at first sight the most favorable scenery for EM because they sample the whole stellar surface, the actual situation is more subtle.
On one hand, there is no useful data from the totality phases of the light curve, so the amount of information may be less than with partial eclipses.
On the other hand, higher inclinations \textch{-- for which the eclipses become total --} do not necessarily imply a better reconstruction. As the system approaches the edge-on configuration, the eclipses start to sample multiple areas in exactly the same way. 
This latter is illustrated in Fig.~\ref{fig:egeom}. The ingress and egress arcs, corresponding to the companion's projected limb on the stellar disc at various orbital phases, form a 'sampling grid'. Each pixel of the grid is eclipsed at one phase interval and reappears at another, contributing to exactly two data points of the differential light curve, interpreted in terms of flux changes from phase to phase%
\footnote{In fact, reconstructing on the sampling grid would be most welcome, because its elements hold the discrete 'packets' that are coded into the light curve. Unfortunately, the pixel areas vary significantly on such a grid; moreover, the grid depends on the sampling of the observed dataset, which may change from eclipse to eclipse -- although the latter would be easily overcome by resampling the data on a uniform temporal grid.}%
.
The left panel shows a configuration in which all the ingress and egress arcs intersect each other in at most one point. The right panel in turn presents a \textch{configuration} in which \textch{most} arcs have two intersection points, so there are pairs of pixels that are eclipsed simultaneously and reappear simultaneously. No inversion method will be able to separate their contribution from each other. 

It is easy to see that these equivalent pixel pairs are arranged symmetrically with respect to the trajectory of the secondary's centre, projected on the sky.
If the trajectory intersects the primary's disc, then there will be a horizontal band similar to that of the uneclipsed region, full of equivalent pixel pairs, and therefore it is an \textit{ambiguous region}, for which the reconstruction is likely to be distorted.
The remaining area in between represents the \textit{trusted region}, for which the unique sampling makes a trustworthy reconstruction possible.

The vertical extents of the limiting regions can be easily computed for spherical stars. They are also approximately valid for real stellar shapes, if polar radii are used. 
Thus, for stellar radii $R_1$ and $R_2$, binary separation $a$, and inclination $i$,
the uneclipsed region has a fractional height  $h_{\rmn{V}} = 1/2 - (R_2-a\cos{i})/(2R_1)$, 
while for the ambiguous region it is
$h_{\rmn{A}} = 1-a\cos{i}/R_1$, all in units of the stellar diameter.
For inclinations larger than $i_{\rmn{min}} = \arccos((R_2-R_1)/a)$, the eclipses are total (formally $h_{\rmn{V}}\!\!<\!\!0$). The ambiguous region disappears for inclinations smaller than 
$i_{\rmn{max}} = \arccos(R_1/a)$. Inclinations between these two limits correspond to the best cases, in which the whole stellar disc is unambiguously sampled. 
However, the existence of such an optimal range requires that
$i_{\rmn{max}} \ge i_{\rmn{min}}$, which implies $R_2 \ge 2R_1$; that is, 
the secondary must be at least twice as large as the primary. Although such systems
may exist, in most cases the secondary is not so much larger. In addition, not all systems show total eclipses. The reconstruction therefore will always be compromised to some extent by the limits of the eclipse geometry. But since partial eclipses may not be as bad compared to total eclipses as it may seem at first, any inclination for which the whole eclipsed region is uniquely sampled is close to optimal.

If the eclipses are central ($i=90\degr$), Eclipse Mapping will only reconstruct modes that are symmetric with respect to the orbital plane. In an aligned rotator, asymmetric modes ($\ell -m = \rmn{odd}$) are subject to complete cancellation \citep{chadid.etal01}, which also persists during the eclipses due to their symmetry, so only symmetric modes are detected anyway. However, in an oblique rotator (not pulsator!), non-symmetric modes will be sampled in a symmetric way, and therefore will not be properly reconstructed. Such edge-on systems are expected to be rare, though.

From Fig.~\ref{fig:egeom} it is also clear that the resolution of the sampling decreases in the vertical direction, so the southern regions (from the observer's point of view) will be reconstructed with poorer quality. There is no such intrinsically uneven sampling in the horizontal direction. Hence, for an aligned rotator, the phase profile will be reconstructed reasonably well, which allows a reliable identification of $m$; but the amplitude profile will suffer from imperfections in the southern parts. 
Fortunately, \textch{for the rotationally symmetric pulsations appropriate for the method} it suffices to know the pattern on one \textch{half} of the stellar hemisphere in order to identify the modes, so the eclipse geometry might not have a dramatic impact on the mode identification.
On the other hand, an oblique rotator has more subtle implications, and the above arguments are not applicable; both profiles will be affected to some extent by the uneven vertical sampling.

To investigate how the limitations discussed above affect the reconstructions, we have selected four system configurations for testing (two distinct systems viewed under two different inclination angles each), listed in Table~\ref{table:systems}, and representing various levels of optimality, as indicated by the heights of the key regions. 
System~1 is an ideal system, with the secondary more than twice as large as the primary and an inclination which enables unambiguous sampling of the whole surface and makes the eclipses total. System~2 is the same system seen at a higher inclination, with an ambiguous region. 
System~3 features both an uneclipsed and an ambiguous region, while System~4 has almost total eclipses but an ambiguous region of considerable area.
\begin{table}
\begin{center}
 \caption{Binary system configurations selected for reconstruction tests.}
 \label{table:systems}
 \begin{tabular}{lllllllll}
  \hline
   System & $R_1$ & $R_2$ & $i$ & $h_{\rmn{V}}$ & $h_{\rmn{A}}$ & $h_{\rmn{tr.}}$ \\
  \hline
   1 & 0.153 & 0.352 & 79.4 & 0    & 0    & 1      \\
   2 & 0.153 & 0.352 & 83.6 & 0    & 0.24 & 0.76   \\
   3 & 0.195 & 0.265 & 80.7 & 0.23 & 0.17 & 0.60   \\
   4 & 0.195 & 0.265 & 85.1 & 0.04 & 0.56 & 0.40   \\
  \hline
 \end{tabular}
\end{center}
   \medskip 
   $R_1,R_2$ are stellar radii in units of the binary separation. 
   $h_{\rmn{V}}$ and $h_{\rmn{A}}$ are the fractional heights of the uneclipsed
   and ambiguous regions respectively; the remaining amount is the trusted region with height $h_{\rmn{tr.}}$.
\end{table}
%

%
\section{Testing the Dynamic Eclipse Mapping\label{sec:em-tests}}
%

In order to asses the performance of the method, we have \textch{subjected} it to extensive numerical testing on artificially generated data. The model uses a simple binary geometry, with rigid spherical stars on a circular orbit. 
\begin{textchange}
Likewise, we employ a simple model for the stellar atmosphere: bolometric limb 
darkening, with coefficients taken from \citet{claret.00:ldcof} for an average delta 
Scuti pulsator with $T=7500$~K, $\log g=2.5$, solar composition, and no turbulent 
velocity. In most runs, however, only the linear limb darkening law was used, with the 
coefficient $x_1=0.56$ drawn from the linear part of the polynomial relation.
\end{textchange}

For strictly spherical stars the mutual occultations can be analytically computed. For maximum precision, each pixel is composed of two flat, triangular tiles, where all the quantities (including partial visibilities due to the occultations) are computed separately and then summarized for the whole pixel. 

To avoid the so-called \textit{inverse crimes}, a situation which occurs when
the data generation and the inversion are made with exactly the same setup, thus ignoring the modeling errors and leading to an over-optimistic assessment of the method's performance \citep{kaipio.somersalo.05}, we used two different grids for the direct and inverse parts. 
For generating the artificial light curves, a uniform spherical grid aligned with the rotation axis was set up on the stellar surface, with an exaggerated resolution of $240\times120$ grid elements.
In turn, the reconstruction was done on an adaptive polar grid, applied on the visible stellar hemisphere, with its $z$ axis pointing towards the observer (as discussed in Section~\ref{sec:assumptions}, the reconstructions must be done on the fixed visible hemisphere). 
\begin{textchange}
The grid was chosen so that its projection on the sky consists of concentric rings of equal widths, each ring being divided into a number of identic pixels determined by the condition that their contribution factors to the integrated flux, 
wis., $\mbox{\textit{pixel area}} \times \cos\gamma \times L(\cos\gamma)$
(where $\gamma$ is the aspect angle and $L$ is the limb darkening),
should be as uniform as possible across the grid. 
With this particular choice, the average linear dimension of the pixels 
is also of the same order, and can be easily matched to the average resolution of the sampling grid
(Sec.~\ref{sec:egeom}), ensuring that the quality of the reconstruction is not limited by the pixelization.
In addition, the approximate constancy of the above contribution factors means that the reconstruction of \textit{surface intensities} and \textit{projected fluxes} of the pixels are practically equivalent, and in fact the latter case, being technically simpler, has been implemented in our algorithm.
\end{textchange}

Throughout the reconstructions  we used a grid composed of $30$ such rings, summing up to $\sim3000$ pixels covering the stellar disc.

As demonstrated earlier in \citet{biro.nuspl:05}, in principle it is possible to reconstruct the static equilibrium intensity map along with the pulsations. In practice, however, the appropriate manner is to account for any static flux component during the binary model fitting; \textch{otherwise} it would probably spoil the binary parameters. (The only purpose of re-mapping it with EM would be \textch{during} an iterative refinement of the parameters.) Therefore we consider it as a nuisance factor and leave it out from the current analysis.

We generated evenly sampled synthetic light curves covering a range slightly larger than the eclipses. With typical pulsation frequencies being $10-100$ times the orbital frequency, we chose a sampling interval of $0.00163$, in units of orbital phase. One cycle of the fastest pulsation is thus covered by about $6$ data points. With typical system configurations this amounts to about $150$ data points per eclipse.
Up to $20$ eclipses from successive orbital cycles were involved. In most cases, though, a much smaller number (sometimes even a single eclipse) already yielded a successful reconstruction. 

The observational errors were modelled by adding artificial Gaussian noise of a specified level to the synthesized light curve. 
\begin{textchange}
In this study we use a \textit{detection} signal to noise ratio (snr), the signal being not the total flux, but rather the semi-amplitude of the weakest pulsation mode, as measured in the total flux outside of the eclipses, where it is free of distortions. The noise is $3\sigma$, as usual ($\sigma$ is the parameter of the Gaussian noise). Typically we used  values of $\rmn{snr}=2-10$.
\end{textchange}

\subsection{Mode identification}

\begin{textchange}

\begin{figure}
\begin{center}
\includegraphics[width=64mm]{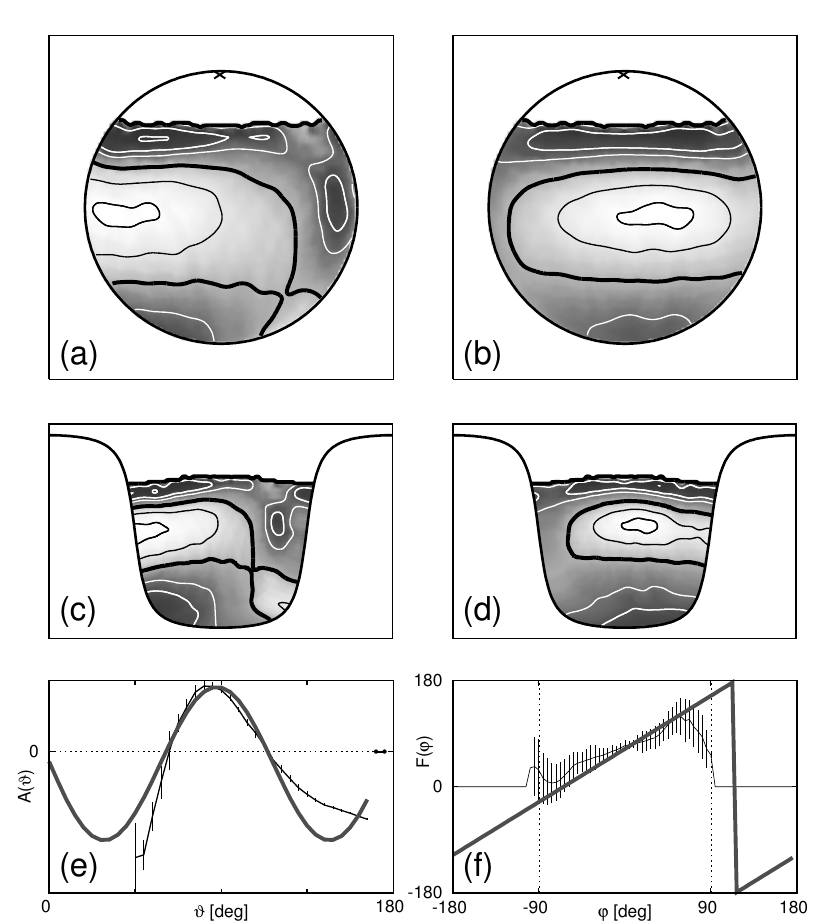} 
\end{center}
\caption{
Illustration of the mode identification procedure.
Panels (a) and (b): the 'cosine' and 'sine' maps of the
reconstructed projected stellar disc for a single mode 
(here $\ell=3$ and $m=1$); crosses mark the position of the rotation
axis.
Panels (c) and (d): the equivalent Mercator maps in the spherical
system of the rotation axis.
Isocontour lines are used to enhance the structure on the 
grayscale maps; thin black and white isocontours mark 
positive and negative values, respectively. The zero isocontours,
corresponding to the node lines, are drawn with thick black lines.
The white "polar caps" are due to the uneclipsed region.
Panels (e) and (f): the amplitude and phase diagrams fitted to
the maps (c) and (d). 
Thick solid lines are the profiles of the input model,
those of the reconstruction appear as thin solid lines, with
the error estimates of the individual points drawn with vertical 
lines.
Dashed vertical lines in panel (f) show the longitude range
corresponding approximately to the visible stellar hemisphere.
\label{fig:mode-ident-example}
}
\end{figure}

\end{textchange}
\begin{textchange}
For each pulsation mode, Dynamic Eclipse Mapping reconstructs a pair of intensity patterns on the adaptive grid applied on the visible stellar hemisphere. They are shown in
panels (a) and (b) of Fig.~\ref{fig:mode-ident-example}, with the addition of limb darkening for visualization purposes -- to look exactly as a close observer would see them in the absence of the other modes, at a given moment $t=0$ and a quarter of period later, respectively.
These data are meaningful after transformed into the spherical coordinate system of the stellar rotation axis (be it given or assumed), where they can be visualized in the usual form of Mercator maps, as illustrated in panels (c) and (d)%
. These maps have the special property that their half corresponding to the invisible stellar hemisphere is empty
\footnote{Although the invisible hemisphere could be interpolated from the visible hemisphere once the mode was identified, that would be confusing and unrealistic.}%
.
Recall that the reconstructions are made on a steady hemisphere, the rotation being redeemed by symmetry assumptions. Eventual uneclipsed regions, lacking any spatial information on the pulsation pattern, will appear also as empty ''polar caps'' near the northern pole of the \textit{orbital} axis, their fluxes being accounted for by virtual pixels.
\end{textchange}

\begin{textchange}
In most cases the pulsation mode can already be guessed by visual inspection of the node lines (isocontours of zero value). However, their identification can be put on more quantitative grounds, by fitting the reference map model (\ref{eq:ref-model}) to the final solution in order to get amplitude and initial-phase profiles, shown in panels (e) and (f)%
. 
(Due to the missing base intensity map, the pulsation patterns have no absolute scale, and that is why only the zero level is marked on the amplitude profile diagrams.)
The slope of the phase profile in the central parts of the stellar disc ($|\varphi|\le60\degr$, say), rounded to the nearest integer, gives the azimuthal mode number $m$, and at the same time imposes a lower limit on $\ell$.
Then checking the number and positions of the nodal points (roots) of the amplitude profile within the trusted region, and comparing them with those of the possible associated Legendre polynomials $P_{\ell}^{m}(\cos\theta)$, gives the most probable $\ell \ge |m|$ value.
We followed this simple procedure in determining the mode numbers in each case.

The underlying non-linear fitting algorithm used for fitting the profiles (Levenberg--Marquardt, Sec.~\ref{sec:refmodel}) also furnishes error estimates for the amplitude and phase profiles. However, we found that sample statistics on columns and rows of the polar grid used for their computation give more reliable estimates, therefore we used the latter for the errors.
\end{textchange}

\subsection{Single modes}

\begin{figure}
\includegraphics[width=84mm]{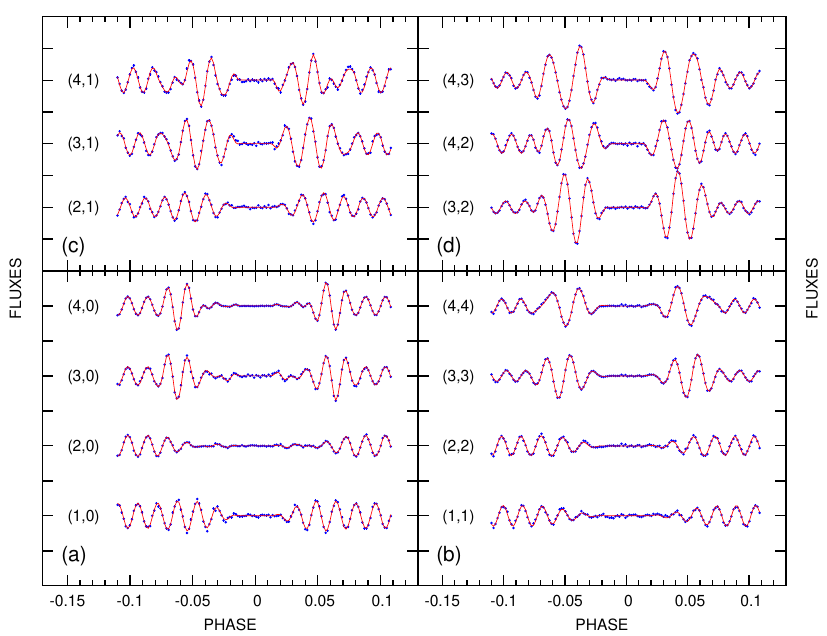} 
\caption{
      The modulation of the pulsations by the eclipses in System~1: single-eclipse 
      excerpts from \textch{noiseless synthetic data}. Panels (a) and (b) summarize the zonal and sectoral modes; the tesseral modes are shown in panels (c) and (d). The mode numbers are shown on the left of the light curves. The fittings achieved by Eclipse Mapping are drawn with solid lines.
\label{fig:lc-samples}
}
\end{figure}

We performed reconstructions of single pulsation modes up to $\ell=4$ on an aligned rotator in all the four systems listed in Table~\ref{table:systems}. 
A frequency of $63.1234$ cycles per orbital period and randomly chosen initial phases were used. Artificial data covered the phases between $\pm0.11$ around 1 to 5 successive eclipses, and had a \textch{\it detection} signal to noise ratio of $10$. 
Noiseless sample modulation light curves in System~1 are shown in Fig.~\ref{fig:lc-samples}. The orbital phases affected by the eclipse are approximately between $\pm0.082$; as it can be seen, the eclipses are total for $|\phi_{\rmn{orb}}| \le 0.02$.

In most cases, data from a single eclipse were enough for a reliable reconstruction, although a few of the modes -- notably the zonal modes with $\ell\ge2$ -- required at least 5 eclipses. Therefore all the cases were investigated with this \textch{eclipse coverage}.

\begin{figure*}
 \hfill
 \includegraphics[width=16.8cm]{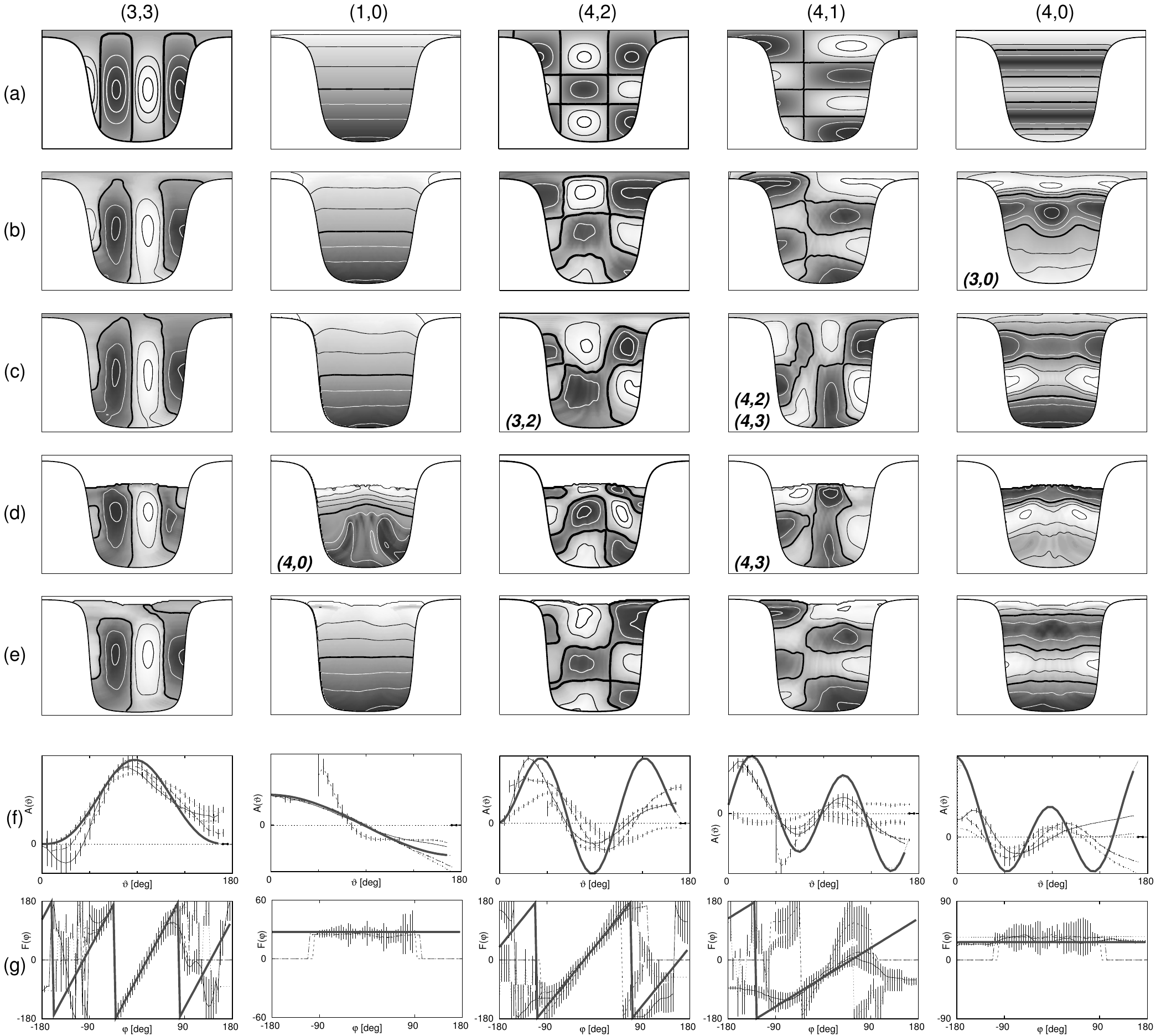} 
 \hfill
\begin{textchange}
\caption{A selection of reconstructed single modes in Systems~1--4, aligned rotator
case.
Each column contains data for a particular mode, marked at the top of the columns,
and ordered from left to right in increasing complexity measure of their vertical structure, $\ell -|m|$ (see text). 
The 'cosine' maps are shown for each case, with contour lines overlaid, as in Fig.~\ref{fig:mode-ident-example}. 
Row (a) shows the input model, rows (b) to (e) contain the reconstructions in Systems 1 to 4. Cases with unsuccessful mode recovery are labelled by the misidentified mode numbers.
The lowest two rows (f) and (g) show comprehensive amplitude and phase profiles for all systems. In these diagrams, thick dashed lines correspond to the model profiles,
while the reconstructed ones are drawn with thin solid, long dashed, short dashed and dot-dashed lines for Systems 1--4, respectively. 
\label{fig:rec-single}
}
\end{textchange}
\end{figure*}

\begin{textchange}
Fig.~\ref{fig:rec-single} shows the reconstructions for a selection of modes.
\footnote{The full set of reconstructions, including modes with negative $m$, is available
in the online material.}%
.
\end{textchange}

For completeness, we \textch{repeated some of the reconstructions using other frequencies}, about three times lower and higher than our adopted value, respectively. The results were the same in all cases, meaning that the pulsation frequency does not play a significant role in the reconstruction, as long as the pulsation is adequately sampled.

\textch{In general, the features of the reconstructions} are in agreement with the limitations discussed in Sec.~\ref{sec:egeom}. The phase profiles are uniformly restored over the central part of the stellar disc in almost all cases, as clearly demonstrated by the amplitude and phase diagrams of Fig.~\ref{fig:rec-single}: for longitudes within $\pm60\degr$, almost all the phase profiles run together with the model. There is only one exception: mode $(4,1)$ in Systems~2 and 3. 
For the amplitude profiles, the \textch{reliability} of the reconstruction varies with the latitude as predicted; the northern half is far better reproduced than the southern. In addition, a peculiar reversal of the patterns may be observed near the north pole, leading to an 'overshooting' in the amplitude profiles, which causes false nodes near 
\textch{$\vartheta=0\degr$}. This phenomenon appears when the input models have a node at the pole; it is absent from the zonal modes. Fortunately the photometrically detectable modes with $\ell\le 4$ do not have nodal lines so close to the polar regions, so this artefact should cause no confusion.

\begin{textchange} For each investigated system there were a few cases for which the reconstruction yielded false mode numbers, at least with the default setup of $5$ eclipses. We list them in Table~\ref{table:missed-modes}. 
While System~1 is indeed the best of all as expected, the differences are not large.
It seems that neither the large uneclipsed regions (as in System~3) nor the ambiguous region (as in System~\textch{4}) cause any dramatic degradation of the profiles. Our conjecture is that \textch{the} specific reference map updating scheme has a significant role in this rather nice behavior.
\begin{table}
\caption{List of \textit{mis}identified modes in the test systems}
\label{table:missed-modes}
\begin{center}
\begin{tabular}{ccc}
\hline
System & original & reconstructed as\dots \\
\hline
1      & $(4,0)$ & $(3,0)$ \\
\hline
2      & $(3,0)$ & $(4,0)$ \\
       & $(3,1)$ & $(2,1)$ \\
       & $(4,1)$ & $(4,2)$ or $(4,3)$ \\
       & $(4,2)$ & $(3,2)$ \\
\hline
3      & $(1,0)$ & $(4,0)$ \\
       & $(4,1)$ & $(4,3)$ \\
\hline
4      & $(3,0)$ & $(4,0)$  \\
       & $(4,0)$ & $(3,0)$ or $(4,0)$\\
\hline
\end{tabular}
\end{center}
\end{table}
\end{textchange}

\begin{textchange}
Looking at Table~\ref{table:missed-modes}, one may observe a decreasing quality of the reconstruction with increasing $\ell-|m|$. 
\end{textchange}
This difference is the number of latitudinal nodal lines in the amplitude profiles (not counting the polar nodes), and is therefore a rough measure of the pattern complexity in the latitudinal direction. 
\begin{textchange}
However, even the most complex mode, $(4,0)$, could be recovered in 3 out of 4 systems.
%
\end{textchange}

In general, it can be concluded that the success of mode \textch{identification} depends primarily on the mode's complexity in the direction perpendicular to the sampling direction of the eclipses. That said, \textch{in the case of a primary with the rotation axis} tilted by $90\degr$ sideways (that is, in the plane of the sky), for instance, zonal modes would be the most favorable cases, while the algorithm would have difficulty in restoring mode $(4,4)$, for example. 

\subsection{Towards reality: multimode pulsations and oblique rotators}
\label{sec:multimode}

\begin{textchange}
 Single mode nonradial pulsation is of course an ideal case. Real pulsators have at least a dozen simultaneous modes. 
Given the frequencies of the detected modes, however, Eclipse Mapping is able in principle to separate each mode's contribution from the others.
\end{textchange}
\begin{table}
\begin{center}
\caption{Parameters of the simulated three-mode pulsational case. The amplitudes
are in arbitrary units and refer to the semi-amplitude in the integrated flux outside the eclipses.}
\label{table:3mode-params}
 \begin{tabular}{clcr}
 \hline \\
 Mode & \multicolumn{1}{c}{Frequency} & Amplitude & \multicolumn{1}{c}{Initial phase} \\
      & \multicolumn{1}{c}{$[\omega_\rmn{orb}]$} &  & \multicolumn{1}{c}{$[\degr]$} \\
 \hline \\
 $(1,0)$ & 59.153517  & 1.387 &  143.56 \\
 $(3,1)$ & 61.547029  & 1.000 &   65.44 \\
 $(2,2)$ & 65.787702  & 1.839 &   12.05 \\
\hline \\
\end{tabular}
\end{center}
\end{table}
\begin{figure}
 \includegraphics[width=8.4cm]{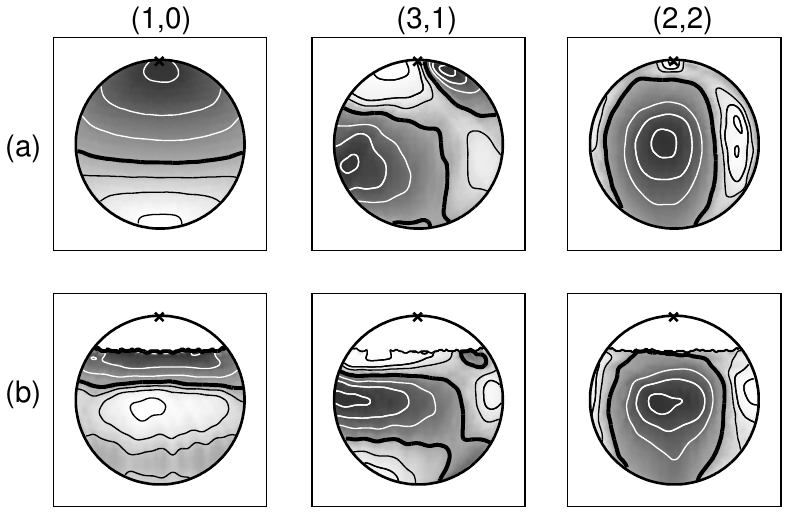} 
 \caption{
  Eclipse maps of a triple-mode pulsation of an aligned rotator in systems 1 and 3 (rows 'a' and 'b', respectively). Mode numbers are indicated at the top of each column. 
  Shown are the 'cosine' maps of the projected stellar disc.
  The $\times$ symbols mark the approximate position of the rotation axis. }
\label{fig:mmA-maps}
\end{figure}
\begin{figure}
 \includegraphics[width=8.4cm]{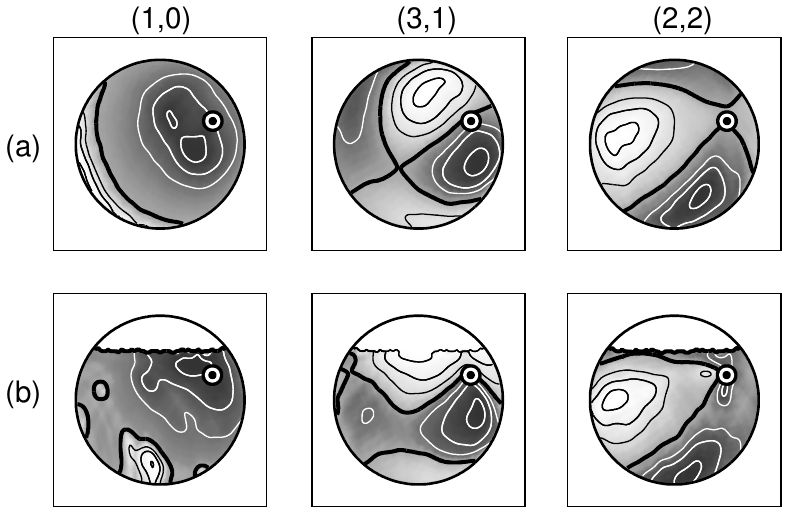} 
 \caption{
  Eclipse maps of a triple-mode pulsation of an oblique rotator in systems 1 and 3 (rows 'a' and 'b', respectively). Mode numbers are indicated at the top of each column. 
  Shown are the 'cosine' maps of the projected stellar disc.
  The $\sun$ symbols mark the approximate position of the rotation axis. }
 \label{fig:mmB-maps}
\end{figure}
\begin{textchange}
Fig.~\ref{fig:mmA-maps} presents a test case of three simultaneous modes with parameters
listed in Table~\ref{table:3mode-params}, and reconstructed on an aligned rotator in Systems 1 and 3. (Sky-projected images are shown in order to allow a comparison with the oblique rotator case below). 
The frequencies were intentionally chosen close to each other, to see how well their modes can be separated by the method. We also decreased the signal to noise ratio to $5$. 
Accordingly, $10$ eclipse cycles had to be included for a good reconstruction.

Similarly, in real systems the axis of rotation is not necessarily aligned with the orbital axis, especially for wider binaries with weak tidal interactions. In this case, however, the direction of the rotation axis must be known a priori.
%
Assuming that this is the case, the specific reference map scheme provides a reliable
reconstruction for tilted axis as well. Fig.~\ref{fig:mmB-maps} shows the repetition of the triple-mode pulsation case for a rotation axis tilted with Eulerian angles
$(\phi,\theta,\psi) = (41\degr,63\degr,0\degr)$  -- that is, rotate the meridian around the initial axis by $41$ degrees, then tilt the axis in the new meridional plane $63$ degrees towards the equator. The third rotation around the new axis would be equivalent to a shift in the initial phases of the pulsation patterns, and was therefore omitted. As a side effect, the longitudes corresponding to the visible disc shift from $[-90,90]$ to about $[-180,0]$. 

Based on the topology of the node lines, it can be seen that the mode numbers  $(1,0),(3,1)$ and $(2,2)$ are successfully recovered in both cases. 
This is also confirmed by the amplitude and phase profiles (available in the online material).
\end{textchange} 

\goodbreak
\subsection{The role of multiple eclipses}
\begin{table}
\caption{Frequency sets for multimode reconstructions.}
\label{table:3mode-freqlist}
\begin{tabular}{llll}
 \hline 
 \multicolumn{1}{c}{Set} & $\omega_1$ & $\omega_2$ & $\omega_3$ \\
 \hline
 1  &  59.153517  & 61.547029  & 65.787702   \\  
 2  &  59.0       & 61.0       & 63.0        \\  
 3  &  59.153517  & 61.153517  & 63.153517   \\  
\hline
\end{tabular}
\end{table}

\begin{textchange}
If the frequencies are not in resonance either with the orbital motion (as would be the case for tidally induced oscillations) or with each other (as would occur for mode interaction), then, \textch{for every mode,} the surface patterns at a given orbital phase will differ from eclipse to eclipse; therefore every new observed eclipse \textch{will present independent pieces of information to the algorithm,
leading to a better separation of the modes, thus improving the quality
of the reconstruction.}
\textch{The increasing amount of independent information will also  compensate for the lower quality of a noisy dataset.}
%

Conversely, for tidally excited pulsations, where the frequencies are multiples 
of the orbital frequency, the flux modulation repeats itself exactly from orbit to orbit, making the additional eclipses largely redundant. 
A similar scenario is the frequency splitting of 
non-radial modes of a synchronously rotating primary, where the differences of the frequencies are multiples of the orbital frequency. Although the rotational splitting is not strictly uniform due to additional second- and higher order factors \citep{soufi.etal.1998:rot.puls,goupil.2000:rot.puls.dsct}, nevertheless they will be quite close to resonance, and pose the same problem, this time the repetition of the relative phase differences between the split modes.
In these cases, multiple eclipses would not be able to improve the reconstruction above a certain limit. Their only benefit will be the improvement of the observational signal-to-noise ratio.
\end{textchange}

To investigate the above possibilities, we repeated the oblique rotator case for System~1 (to keep the complexity level reached so far) with two other frequency sets, \textch{simulating the two resonance possibilities discussed above}. They are listed in Table~\ref{table:3mode-freqlist}, with the original one in the first line.

\begin{textchange}
\begin{figure*}
 \includegraphics[width=16.8cm]{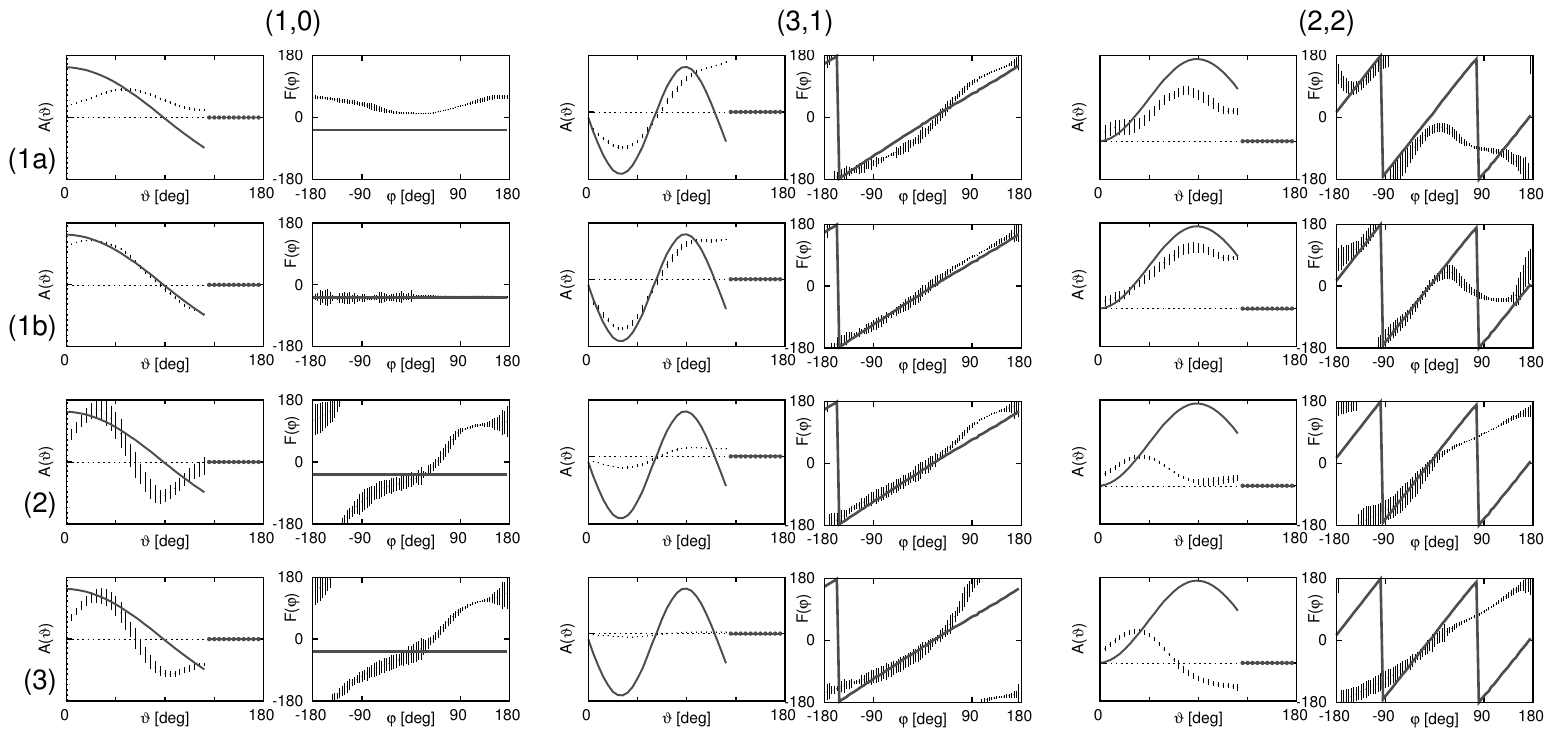} 
 \caption{The dependence of the reconstructed profiles on the number of involved eclipses.
          Amplitude and phase diagrams are shown for each of the 3 modes.
          Reconstruction was made in the oblique rotator of System~1.
          Rows contain the following: (a) -- non-resonant frequencies, 2 eclipse cycles;
          (b) -- non-resonant frequencies, 10 cycles; (c) -- tidally resonant frequencies,
           10 cycles; (d) -- uniformly splitted frequencies, 10 cycles.
      }
\label{fig:multimode-multiecl}
\end{figure*}
The results are compared in terms of the reconstructed profiles in Fig.~\ref{fig:multimode-multiecl}.
In the non-resonant case, two eclipses still give poor results, in particular for the first mode (row (a)), but raising the number of eclipses to $10$ (row (b)) provides enough information for a proper mode identification.
In contrast, the results for the resonant cases, shown in rows (c) and (d) for the same number of $10$ eclipses, demonstrate that the improvement of the reconstruction expected from the inclusion of additional eclipses is prevented by the resonances. In our particular case, the fitted profiles lead to the estimates $(3,1)$, $(3,1)$, and $(4,2)$, instead of the true figures $(1,0)$, $(3,1)$, and $(2,2)$.
\end{textchange}

Tidally induced oscillations and synchronous rotation occur primarily in close binaries. The mapping in such systems is already complicated by other proximity factors (tidal distortion effects, mass transfer), and the limitation discussed above is one more reason why wider binary systems are more preferred by Dynamic Eclipse Mapping. 
It should be emphasized, though, that once the effects of interaction are properly accounted for, then the close binaries with non-resonant pulsations become more attractive, because the components are more likely to contain aligned rotators 
\textch{-- that is, with a known direction of the spin axis}.

\subsection{Errors in the modelling parameters}

All the tests made so far assumed \textch{the true} values for the physical parameters. Since EM by its nature is sensitive to the accuracy of the data, it is sensible to ask what impact do inexact \textit{model parameters} have on the results. Parameters that may affect the reconstructions are: the frequencies of the pulsation, the geometric parameters (stellar radii, inclination, \textch{and the direction of the spin axis}), and the limb darkening.

Recent space-based, continuous, long-term observations allow a very precise determination of the pulsation frequencies, therefore serious errors in the frequencies shouldn't happen these days.

Likewise, the geometric parameters can in general be determined to a high accuracy in eclipsing binaries. 
However, there are well-known correlations between the stellar radii and the inclination, in that about the same eclipse profile may be reproduced by different combinations of their values. With poor or missing spectroscopic data, more parameters have to be 'guessed', and the correlations might lead to biased parameters.
\begin{textchange}
The problem of incorrect or unknown direction of the rotation axis is more complex, and is left for a subsequent paper.
\end{textchange}

To investigate the effects of correlated deviations from the true values, we created three sets of binary parameters by shifting the inclination by $3\degr$ in both directions from a central value, and modifying the stellar radii accordingly by a trial and error method, to get about the same primary eclipse profile. The parameters are listed in Table~\ref{table:3mode-tuned}. The limb darkening was kept fixed.
We then generated data for a triple mode pulsation in the middle system with  $i=82.4\degr$. We \textch{used} the same pulsation model as in 
Table~\ref{table:3mode-params}, with the difference that the third mode $(2,2)$ was replaced by $(4,2)$, the latter being more complex and therefore more liable to errors.
The same dataset was then reconstructed with all three parameter sets.
\begin{table}
\caption{List of three equivalent binary parameters giving similar primary eclipse profiles.}
\label{table:3mode-tuned}
\begin{tabular}{llll}
 \hline 
 Set & $R_1$ & $R_2$ & $i$ \\
     & [sep] & [sep] & [deg] \\
 \hline 
  1  & 0.153 & 0.352 & 79.4 \\
  2  & 0.160 & 0.325 & 82.4 \\
  3  & 0.173 & 0.302 & 85.4 \\
 \hline 
\end{tabular}
\end{table}

\begin{figure*}
 \includegraphics[width=16.8cm]{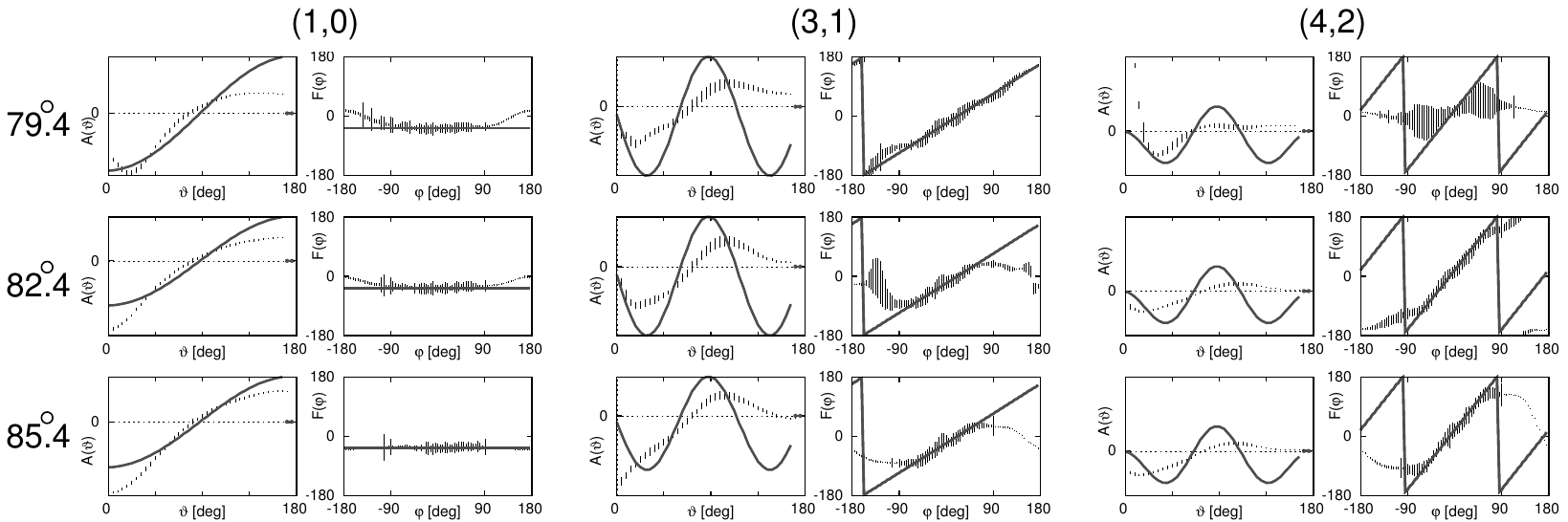} 
\begin{textchange}
\caption{The effect of biased geometric parameters on the reconstruction, shown in terms of amplitude and phase profiles. Each row shows results for a particular inclination (shown at left). Other notations are the same as in Fig.~\ref{fig:multimode-multiecl}.
 \label{fig:3mode-tuned-rec}
}
\end{textchange}
\end{figure*}

The results are \textch{summarized} in Fig.~\ref{fig:3mode-tuned-rec}.
The first two modes are more or less restored in all cases. The third mode, being more complex, suffers larger distortions, and is not recovered with the lower inclination, where the slope of the phase profile is $\sim\!0.73$, while the amplitudes are way off. A formal mode identification would give $m=1$ and $\ell=3$, totally erroneous because mode~2 already has this set of ($\ell,m$). There is no such problem with the higher inclination case, though.
We conclude that although small errors in the binary parameters are not a serious obstacle against a successful EM, they may cause a misidentification of the more complex modes.

Another important factor is the limb darkening. The \textch{oscillation amplitude caused by a given mode} in the whole-disc integrated flux depends significantly on the adopted limb darkening law, and so does the \textch{magnitude} of the modulations during the eclipses. All our previous runs have been made with the assumption of a linear limb darkening law. With the full polynomial law, all modes would produce larger distortions during the eclipses -- sometimes by orders of magnitude, as is the case for $\ell=3$ modes, for instance. The overall pattern of the distortions, however, remains the same. The net effect is that EM would give false amplitudes for the maps, but the pattern would not be modified, therefore the mode identification is not hampered by small errors in the limb darkening. 

We did check the above conjecture by additional tests, although we limited ourselves to linear and polynomial limb darkening laws for the \textit{same} stellar model, because it is enough to induce changes near the disc limb -- the place where the differences really matter \textch{for the Eclipse Mapping}. The coefficients were taken from \citet{claret.00:ldcof}. 
Light curves were generated for a couple of selected modes ($\ell=3$ cases inclusive) with polynomial limb darkening, then eclipse mapped with the linear law. The other possible combinations were also performed, and the results confirm the expected behavior. The amplitudes indeed varied greatly, 
\textch{but the nodal points of the amplitude profile and the slope of the phase profile both remained the same}. We do not show them here because a similar case occurs implicitly in the tests of the next section. 

\subsection{Hidden modes}
Due to symmetries in the pulsation patterns, for each mode there are certain axial orientations for which the condition of partial or complete cancellation occurs, that is, the integration over the visible stellar hemisphere gives zero net flux variation. 
\begin{textchange}
For example, all antisymmetric modes (with $l-|m|=\rmn{odd}$) seen \textch{edge-on}, as well as all sectoral modes \textch{seen pole-on}, are subject to cancellation.
For each mode there are also certain intermediate angles at which complete cancellation occurs \citep[see][for an extensive study]{chadid.etal01}. 
\end{textchange}

Now in eclipsing binaries the inclination of the orbit is close to $90\degr$. 
Antisymmetric modes on an  aligned rotator thus will be close to the condition of complete cancellation. They show up during the eclipses however, because the symmetry of the surface integration that led to the cancellation effect is lifted on a partially
\textch{non-centrally} occulted disc \textch{(provided that the system is only close to, but not exactly at the edge-on configuration)}.
Because it is customary in time series \textch{analysis} to ignore the data segments affected by the eclipses so as to avoid the complications (sidelobes and false peaks) caused by modulated sinusoids, such 'hidden modes' may go undetected. Therefore their signal, amplified during the eclipses \textch{and} unaccounted for by any frequency, certainly will affect the \textch{reconstruction}.

\begin{figure}
 \includegraphics[width=8.4cm]{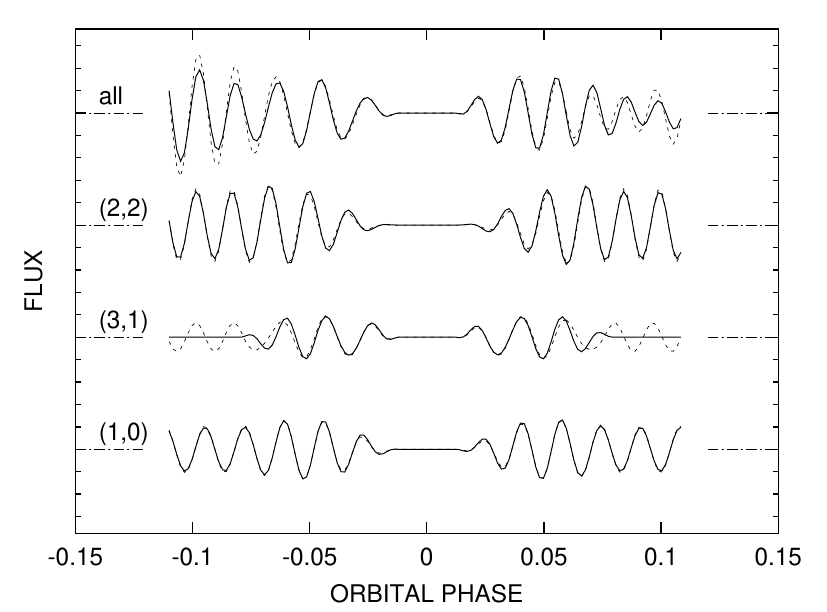} 
 \caption{Simulated light curves for linear (dashed lines) and non-linear (solid lines) limb darkening laws of the same stellar model. The top curves show the overall fluxes, while the others detail the contributions of the individual modes. Each \textch{group} of curves has its base level marked by the dot-dashed segments around them. Note in particular the large differences for mode $(3,1)$.}
 \label{fig:mm3-ldeffect}
\end{figure}

The aligned rotator with multiple modes presented in Section~\ref{sec:multimode} contains in fact such a hidden mode, $(3,1)$. As mentioned in the previous section, $\ell=3$ modes were found to be extremely sensitive on the assumed limb darkening.
So we repeated exactly that case, this time with a non-linear limb darkening.
Fig.~\ref{fig:mm3-ldeffect} compares the generated light curves, also \textch{separately showing} the contributions of each individual \textch{mode}. The second mode, $(3,1)$, is practically hidden, giving almost zero net flux variation outside, 
but a significant contribution during the eclipses. 

\begin{figure*}
 \includegraphics[width=16.8cm]{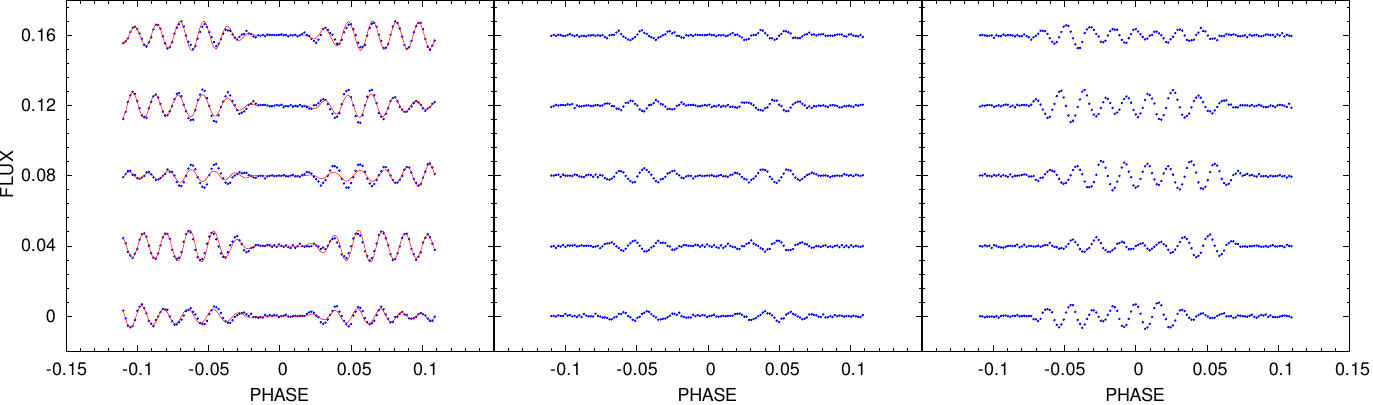} 
\caption{Light curves and residuals for the multimode case with hidden mode. The five eclipse cycles are folded and shifted vertically by $0.04$ from each other. The left panel shows the synthetic light curve and the fit achieved by the EM. The residuals are shown in the middle panel. Compare this especially with the modulations of the third mode, shown in Fig.~\ref{fig:mm3-ldeffect}. The right panel shows the residuals after the subtraction of the  two-frequency sinusoidal fitted with \textsc{Period04}, for a comparison.
\label{fig:mm3-em-fit-with-two}
}
\end{figure*}

\begin{figure}
\includegraphics[width=8.4cm]{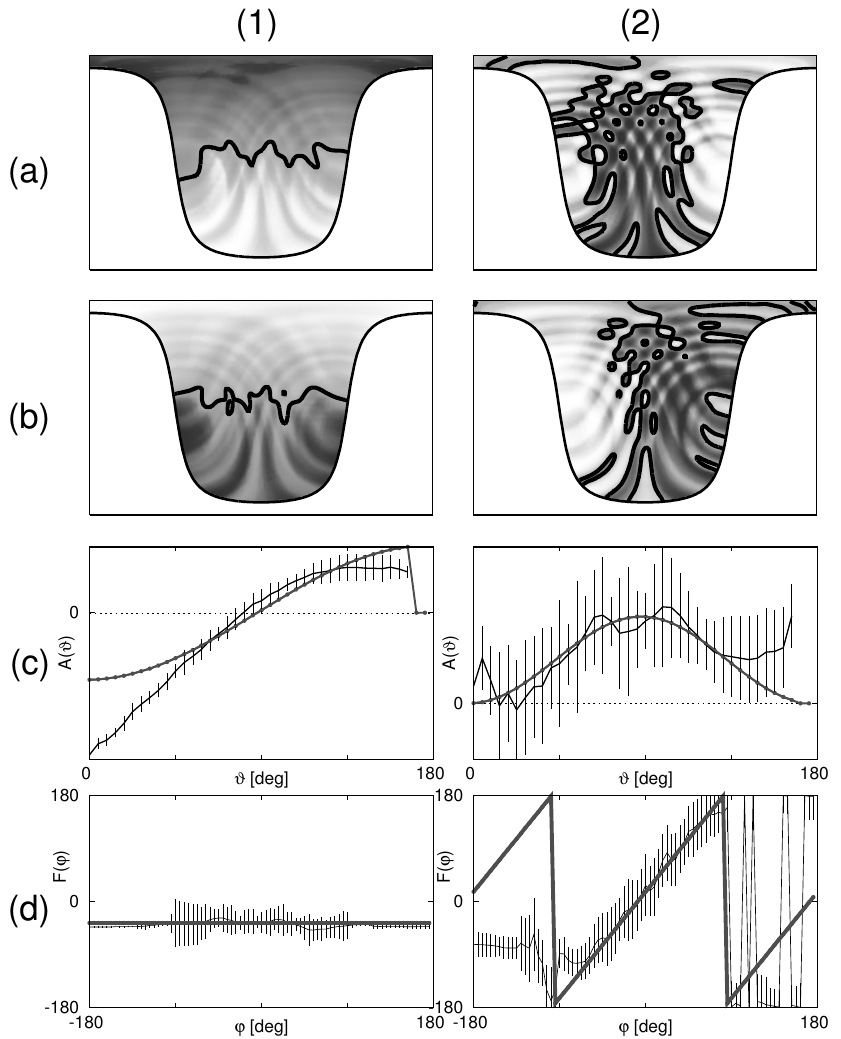}
\caption{Reconstruction of the multimode case with hidden mode. The top row shows the reconstructed map pairs for the two modes, the second and third rows contain the amplitude and phase profiles of the solution, and in the bottom row we show the maps constructed from the fitted profiles.
\label{fig:mm3-em-maps-with-two}
}
\end{figure}

To fully simulate the effect of hidden modes, we searched for frequencies in the artificial data rather than using the model values. For this purpose a larger dataset was generated, covering $5$ consecutive eclipses but extending over almost the full orbital cycle, in order to contain enough data for a time series analysis. We used the program \textsc{Period04} \citep{p04.reference} to derive the frequencies, after having cut out the \textch{segments affected by the eclipses}. The dataset could  be \textch{perfectly fitted} with two frequencies $\omega_1=59.1545436$ and $\omega_2=65.7879857$, which are very close to the input values (lines 1 and 3 in Table~\ref{table:3mode-params}). 

We then fed these two frequencies and the original dataset to the EM.
The algorithm had trouble in achieving a good fit to the data. The lowest attainable values for the R-statistics were $R\sim128$, implying $\chi^2=5.6$, and gave very messy images. Thus we had to significantly relax the fitting criteria from $R=1$ to $R=150$ for an acceptable solution; the $\chi^2$ also increased accordingly to $6.5$. 
Fig.~\ref{fig:mm3-em-fit-with-two} shows large residuals that the algorithm could not fit, those being due to the hidden mode. That they were not falsely mapped on the other modes is reassuring, in fact. Indeed, the modes could be successfully restored, though with a higher uncertainty, as \textch{shown} in Fig.~\ref{fig:mm3-em-maps-with-two}. 

When we included the eclipses in the time series analysis, even $20$ frequencies were not enough to properly account for the modulations during the eclipses. Two of these frequencies corresponded to modes 1 and 3, and a third frequency was near to that of the hidden mode, but its amplitude was of the same order as \textch{of} other $17$ peaks in the spectrum. \textch{This is not unexpected, since a purely Fourier-based spectral analysis is only appropriate for purely harmonic oscillations}.
On the other hand, subtracting the model fitted by EM from the original dataset leaves us with a much cleaner residual light curve, because now EM properly accounts for the distortion effect of the eclipses on the two modes with known frequencies. A Fourier spectral analysis of these residuals is still cumbersome, with \textsc{Period04} giving two main frequencies  $\sim50.55$ and $60.55$ but still nothing at the location of the true frequency $61.547$.
More specific approaches would be needed for this purpose. We mention one promising technique described in \citet{bretthorst.88:sp.an}, which, \textch{with an appropriately designed model, is expected} to handle the modulations of the amplitude and the instantaneous frequency caused by the eclipses.

Nevertheless, our investigation demonstrates a certain immunity of EM against the pollution by the hidden modes, in that they do not jeopardize the reconstruction of \textch{the} other modes; moreover, EM is able to properly isolate their contribution from the detected modes.
Of course, we are aware that \textch{the} simple approach outlined above may not work as efficiently in real circumstances. For example, multiple hidden modes may produce residuals that could be isolated only with the inclusion of an unrealistically large number of eclipses, \textch{or even worse, could lead to a combination that cannot be isolated from the signal of the detected modes}.

\section{Conclusions}

In this paper we have introduced \textch{the} Dynamic Eclipse Mapping method, designed to reconstruct the pulsation patterns of non-radially oscillating components in eclipsing binaries, with the goal of mode identification. The method uses the effective surface sampling of the eclipses
and provides image-like information on the pulsation patterns, which eventually enables a direct mode identification, without the need to invoke detailed models of stellar structure and atmosphere. Only a geometric model for the binary and a few simple assumptions on the stellar atmosphere are needed. The method takes the detected frequencies and the eclipse light curve as input data, and furnishes pairs of images for each mode, completely describing its spatial and temporal behavior. A particular advantage of the method is that it can in principle operate on any wavelength range, including wide-band photometric data, making the datasets obtained in space observatories potentially useful for more than time series analysis.

We have performed extensive testing of the Dynamic Eclipse Mapping. Based on these tests, we \textch{can} make the following conclusions.
\begin{enumerate}
 \item 
       The reconstructions do not depend dramatically on the eclipse geometry. In particular, partial eclipses are not an obstacle, \textch{provided that their
       contribution is properly treated in terms of regularization and fitting}. We believe that the \textch{use of an} axially symmetric reference map updating scheme \textch{is essential} in providing this rather nice property. Inclinations very close to $90\degr$, however, are less favorable cases, due to the increasing symmetry of the eclipses' surface sampling.
\item 
       \textch{Chances for a successful mode identification decrease with an increasing complexity of the pulsation pattern} in the direction perpendicular to the secondary's projected orbit, measured roughly by $\ell-|m|$. 
       There is no such limit in the horizontal direction.
\item 
       Multimode pulsations can also be reconstructed, provided that the data cover a sufficient number of individual eclipses. Although we \textch{only presented a case with} $3$ simultaneous modes, \textch{there is no} procedural obstacle in reconstructing a larger number of simultaneous pulsations, if the eclipse coverage is large enough to allow a proper separation of all the detected modes. The triple mode case required $5-10$ eclipses, a dozen of modes would certainly need 
       \textch{much more, which calls for uninterrupted space-based observations, and probably needs a lot more computational time as well}.
\item   
        Modes resonant with the orbital motion are problematic in that the inclusion of subsequent eclipses \textch{(after a certain limit, determined by the resonant cycle) will not improve the reconstruction}. For single modes, one cycle may be \textch{enough anyway}, but for multiple \textch{modes} it is certainly a \textch{limitation that no other inversion method would be able to overcome. Tidally excited pulsations fall into this category, as well as the rotational splitting of modes in tidally locked systems, where the difference of the frequencies is in 1:1 resonance with the orbital motion.}
\item  
       Pulsations on an oblique rotator can be reconstructed with similar conditions as in the aligned case, \textch{assuming that the direction of the rotation axis is known}. Without this essential information, EM cannot give any useful results, while wrong assumptions about it will almost surely cause false mode identification.
\item  
       Moderate errors in the geometric parameters are tolerable, although they \textch{may} cause \textch{false identification of the more complex modes}. A correct account for the limb darkening is also important. In particular, it is crucial to go beyond the generic linear limb darkening relation, otherwise the reconstructed amplitudes \textch{may become too distorted}. \textch{This requirement should no cause any difficulty though, with the observational accuracy achievable these days.}
\item  
       The method \textch{tolerates the presence of hidden modes and is able to properly isolate their polluting effect, so that they don't hinder the proper identification of the other modes}.
\end{enumerate}
Based on the above findings, an ideal target for the method would be a moderately wide system, with no significant tidal distortions of the pulsating component(s), no resonant pulsations, yet not as wide as to have an eccentric orbit, so that the rotation axis may be assumed aligned with the orbital axis.
This latter problem may be overcome by high-resolution spectroscopic observations that could reveal the Rossiter--McLaughlin effect and infer a spin axis from it, as it was done for DI~Herculis \citep{albrecht.etal.10:di.her}. The effect has already been detected in RZ~Cas \citep{lm.08:rzcas.rossiter}, and the possibility is open for other cases as well.

Our method does not require that the pulsation patterns be of spherical harmonics type; only rotational symmetry must hold for the modes.
In principle, modes distorted by rapid rotation of the pulsating component can also be investigated. \citet{lignieres.etal.06:papI} have shown that fast rotation causes an equatorial concentration of the pulsation amplitude, while the azimuthal structure is unchanged, so the modes obey the same symmetry as assumed by our method. In addition, it was also shown by the same authors that the number of horizontal surface node lines remains the same, with small shifts in their positions. Therefore the overall topological structure of the modes is unchanged, making them suitable for mapping with EM. 
Moreover, some of the rotationally distorted modes have a larger disc averaging factor than their zero rotation equivalents, making them more easily detectable, and ultimately allowing \textch{the detection of higher degree modes} with $\ell$ up to $6-8$.
Only at extremely high speeds does the equatorial concentration flatten the amplitude profile (except for a small equatorial region) to the extent that any topological information becomes too weak to recognize.
We did not deal with distorted modes in this study, though, because we believe that the importance of such an extension will be settled by the outcome of the first real-world applications.

The simple binary model used here was appropriate for assessing the usability of the Dynamic Eclipse Mapping method. Successful applications obviously will require a more detailed model for the binary, but we \textch{expect} that the inclusion of the neglected features and effects is a computational issue, and should not endanger the success of the mode identification.

\section*{Acknowledgments}
This work was supported by the Hungarian NSF grant OTKA~F-69039 and the
'Lend\"{u}let' Young Researcher program of the Hungarian Academy of Sciences.
The authors wish to thank L\'{a}szl\'{o} L.\/ Kiss and Olivera Latkovi\'{c} for improving the structure of the manuscript with their precious comments.

\label{lastpage}

\end{document}